\documentclass[12pt,journal,final,onecolumn]{IEEEtran}

\usepackage[pdftex]{graphicx}
\usepackage[cmex10]{amsmath}
\usepackage{amssymb}
\usepackage{amsthm}
\usepackage{amsfonts}
\usepackage{bm}
\usepackage{cite}
\usepackage[svgnames]{xcolor} 
\usepackage{pstricks,pst-node,pst-plot,pstricks-add}
\usepackage{hyperref} 

\graphicspath{{figs/}}

\input{niesen.def}

\newcommand{\dtd}{\Delta t_D}
\newcommand{\dta}{\Delta t_A}
\newcommand{\dsd}{\Delta s_D}
\newcommand{\dsa}{\Delta s_A}

\begin{document}

\title{Inter-Vehicle Range Estimation from \\ Periodic Broadcasts}

\author{Urs Niesen, Venkatesan N. Ekambaram, Jubin Jose, Xinzhou Wu%
    \thanks{U. Niesen, J. Jose, and X. Wu are with Qualcomm Research. V. N.
    Ekambaram was with Qualcomm Research; he is now with Intel Research.
    Emails: uniesen@qti.qualcomm.com, venkyne@gmail.com, jjose@qti.qualcomm.com, 
    xinzhouw@qti.qualcomm.com}%
    \thanks{Parts of this paper were presented at the IEEE
        Vehicular Technology Conference, September 2015 \cite{niesen2015}.}%
}

\maketitle

\begin{abstract}
    Dedicated short-range communication (DSRC) enables vehicular communication
    using periodic broadcast messages. We propose to use these periodic
    broadcasts to perform inter-vehicle ranging. Motivated by this scenario, we
    study the general problem of precise range estimation between pairs of
    moving vehicles using periodic broadcasts. Each vehicle has its own
    independent and unsynchronized clock, which can exhibit significant drift
    between consecutive periodic broadcast transmissions. As a consequence, both
    the clock offsets and drifts need to be taken into account in addition to
    the vehicle motion to accurately estimate the vehicle ranges. We develop a
    range estimation algorithm using local polynomial smoothing of the vehicle
    motion. The proposed algorithm can be applied to networks with arbitrary
    number of vehicles and requires no additional message exchanges apart from
    the periodic broadcasts. We validate our algorithm on experimental data and
    show that the performance of the proposed approach is close to that obtained
    using unicast round-trip time ranging. In particular, we are able to achieve
    sub-meter ranging accuracies in vehicular scenarios. Our scheme requires
    additional timestamp information to be transmitted as part of the broadcast
    messages, and we develop a novel timestamp compression algorithm to minimize
    the resulting overhead.
\end{abstract}

\section{Introduction}
\label{sec:intro}

\subsection{Motivation}
\label{sec:intro_motivation}

Intelligent transportation systems have gained significant importance in the
recent past with various initiatives taken both by the federal and state
transportation agencies in collaboration with the automotive industry. Two major
applications in this context are vehicular safety and autonomous driving. A key
requirement for both these applications is precise, sub-meter vehicle
localization \cite{drane1998positioning} typically enabled through ranging
(i.e., distance estimation).  Existing technologies such as LiDAR
\cite{khattak2003application} or RTK-GPS \cite{misra2006global} that provide
such precise ranging have the drawback of being costly to deploy. There is hence
a need for precise ranging solutions that are in addition cost effective and
commercially viable. As a possible solution to this problem, we explore in this
work the use of affordable WiFi hardware for obtaining precise range estimates
between vehicles.

The underlying concept of ranging using radio-frequency signals exploits that
the propagation delay of the wireless signal from the transmitter to the
receiver is proportional to their distance. Time-of-arrival (ToA) ranging
\cite{golden2007sensor} typically assumes that the transmitter and receiver are
time synchronized in order to calculate the propagation delay.
Time-difference-of-arrival (TDoA) ranging \cite{misra2006global} uses a small
number of time-synchronized reference nodes allowing the other nodes in the
system to obtain ranging measurements from these reference nodes without
themselves being time synchronized. Finally, round-trip-time (RTT) ranging
\cite{11mc} eliminates the need for time synchronization by exchanges two
messages between the transmitter and receiver, from which the common clock
offset can be canceled. These two messages need to be exchanged within a short
time interval, since otherwise clock drifts can introduce significant ranging
errors.

Time synchronization is hard to achieve in vehicular applications particularly
when the synchronization accuracy needs to be in the order of nanoseconds to
achieve sub-meter level accuracy. This precludes the use of ToA based ranging
schemes. Similarly, it may not be practical to establish synchronized reference
nodes, which are needed for TDoA based schemes. RTT could be used to obtain
relative range estimates between the vehicles. However, a unicast mechanism like
RTT requires a significant number of message exchanges especially when the number
of vehicles in the network is large. In particular, the number of messages
scales \emph{quadratically} with the number of vehicles.

\subsection{Summary of Results}
\label{sec:intro_summary}

In this work, we develop a broadcast mechanism for range estimation that can
drastically reduce the number of required message exchanges. In particular, in
our proposed scheme, the number of messages being broadcast scales only
\emph{linearly} with the number of vehicles. The broadcast mechanism is motivated
by the dedicated short-range communications (DSRC) standard \cite{dsrc}, which
is a modification of the WiFi standard. The DSRC standard prescribes periodic
broadcasting of safety messages by every vehicle. We propose to use these
broadcasted safety messages to perform ranging without the need for any
additional messaging. Our scheme does require additional information to be
transmitted in each DSRC packet, and we introduce a compression mechanism to
minimize this overhead. 

The contributions of this paper are summarized as follows:

\begin{itemize}
    \item We develop an algorithm and protocol for range estimation between
        vehicles based on broadcast message exchanges wherein each vehicle
        periodically broadcasts a packet with certain timing information. Using
        this periodic broadcast ranging approach, the number of required message
        exchanges scales only linearly with the number of vehicles as opposed to
        quadratically when using standard unicast RTT ranging.
    \item The time interval between broadcast transmissions can be on the order of
        $100$ ms.  As a consequence, we need to explicitly account for the
        relative clock offset and clock drift between the vehicles. Further we
        also need to account for the motion of the vehicles during this time
        interval.
    \item We also propose a novel compression algorithm to efficiently quantize
        and transmit timestamp information in the broadcast packets in order to
        reduce the protocol overhead. This is particularly significant in the
        context of DSRC, in which packets have constrained payload size.
    \item We conduct experiments using Qualcomm-proprietary WiFi hardware that
        supports ranging capabilities. We show that sub-meter level accuracies
        are achievable with the proposed algorithm. Finally, the proposed
        algorithm suffers only a small loss in ranging performance
        as compared to RTT-based ranging. Thus, the reduction from quadratic
        number of messages in RTT to linear number of messages in our proposed
        periodic broadcast ranging algorithm results in only a small loss in
        ranging performance.
\end{itemize}

\subsection{Related Work}
\label{sec:intro_related}

Range estimation has been a topic of research interest for several decades
starting with classical RADAR systems \cite{levanon1988radar} and GPS
\cite{misra2006global} to more recent technologies like WiFi \cite{11mc},
Bluetooth \cite{hossain2007comprehensive}, and ultra-wide band
\cite{lee2002ranging} to name a few. There is still active research in the
community to develop inexpensive and precise ranging technologies that can work
in various environments. These are enabling applications in the vehicular space
\cite{parker2006vehicle,alam2009range,alam2010dynamic,boukerche2008vehicular}, sensor networks \cite{hu2004localization,lanzisera2006rf,denis2006joint}, and robotics
\cite{cao1997cooperative,djugash2006range}. 

GPS works on the TDoA principle wherein the satellites are time synchronized.
RADAR based systems typically work on the principle of electromagnetic
reflection. The transmitted signal is reflected off a target object and the
received reflected signal can be correlated with the transmitted signal to
locate the target. Some high-end cars have RADAR-like sensors that detect
distances to obstacles and other vehicles.  However, when multiple vehicles are
present in the environment it is hard to distinguish between different
reflectors. Further, if all the vehicles transmit RADAR pulses without a
synchronization protocol, then the resulting interference between the signals
can degrade the ranging accuracy.

RTT-based systems are popular and well studied in various contexts. The IEEE
802.11mc task group defines a Fine-Timing-Measurement (FTM) protocol \cite{11mc}
for RTT ranging in the WiFi signal spectrum. As described before, this
eliminates the need for clock synchronization, but is inherently unicast in
nature. There are quite a few works in the literature that consider the problem
of range and location estimation in the presence of unknown clock offsets
\cite{wang2011robust, noh2007novel,zhou2011indoor}. Most of these are in the
context of location estimation, wherein the location of a node or a set of nodes
needs to be estimated from ranging measurements. These works explore the joint
estimation of locations and clock offsets. The problem of joint estimation of
clock parameters and distances between pairs of mobile nodes is considered in
\cite{rajan2012joint}, which makes use of the local smoothness of the node
trajectories to solve for the unknown variables.

\section{System Model}
\label{sec:system}

We consider a collection of moving vehicles, each equipped with a radio
transmitter and receiver, that emit periodic broadcast messages with approximate
period of $T$ seconds. Our goal is to estimate the distances between the
vehicles using time of departure and arrival information from these broadcast
messages. This scenario is motivated by the DSRC standard, in which vehicles
broadcast so-called basic safety messages with an approximate period of $T =
0.1$ s. We will use DSRC as a running example throughout this paper, however the
results apply more generally.

\begin{figure}[htbp]
    \centering 
    \includegraphics{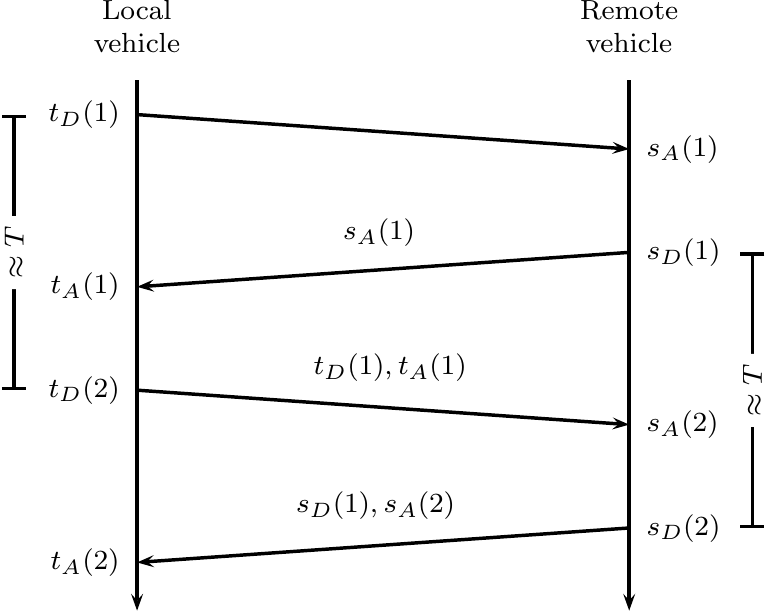} 

    \caption{Message exchanges for ranging using periodic broadcasts.}
    \label{fig:timeline}
\end{figure}

For ease of presentation, we restrict ourselves for now to a scenario with two
vehicles, a local and a remote vehicle. The generalization to arbitrary number
of vehicles is discussed in Section~\ref{sec:main_multiple}. We assume that the
vehicle clocks are not synchronized. Since the message departure and arrival
times are measured with respect to each vehicle's own clock, we need to
explicitly account for the relation between the two clocks. Formally, we denote
by $t_D(n)$ and $s_D(n)$ the departure times of the $n$\/th message sent by the
local and remote vehicle, respectively. Similarly, we denote by $t_A(n)$ and
$s_A(n)$ the arrival times of these messages. These arrival and departure times
are piggybacked on the broadcast messages as illustrated in
Fig.~\ref{fig:timeline}.

\begin{figure}[htbp]
    \centering 
    \includegraphics{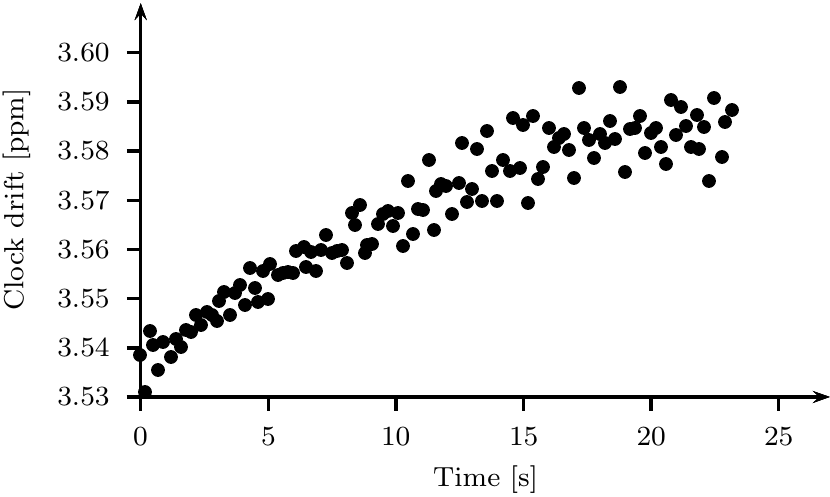} 

    \caption{Experimentally measured WiFi receiver clock drift $\delta$ as a
    function of time. The experiment is carried out in a static setting, and the
    parameter $\delta$ is estimated by differencing consecutive measurements.}

    \label{fig:drift}
\end{figure}

We assume that the local time $t$ and the remote time $s$ are related via the
linear relationship
\begin{equation}
    \label{eq:clock}
    s(t) = \theta + (1+\delta)t,
\end{equation}
where $\theta$ is the clock offset (measured in seconds) and $\delta$ is the
clock drift (a unitless quantity). Typical values of $\delta$ are on the
order of $\pm 10^{-5}$, which is usually expressed as a drift of $10$ parts per
million (ppm); see also Fig.~\ref{fig:drift}. 

In the periodic broadcast ranging approach analyzed in this paper, the delay
between a message departure and the following message arrival is on the order of
$T/2$. Hence, the clock drift during this period is on the order of $\delta
T/2$.  For example, in the DSRC case with $T = 0.1$ s, the drift of a $10$ ppm
clock during that time is on the order of $0.5$ $\mu$s. When left uncompensated,
this clock drift of $0.5$ $\mu$s will result in a ranging error of approximately
$150$ m! 

Careful modeling and dealing with clock drifts is hence critical for successful
ranging using the periodic broadcast approach. Continuing with the DSRC example,
assume we aim for a ranging accuracy on the order of $0.3$ m, then we can
tolerate an uncompensated clock drift of at most $0.01$ ppm between successive
samples.  Fig.~\ref{fig:drift} suggests that we can therefore model the clock
drift as being approximately constant for a window of a few seconds
(corresponding to a few tens of DSRC message exchanges). During this time, the
model~\eqref{eq:clock} with constant $\delta$ is valid. 

The departure and arrival timestamps can then be related as follows:
\begin{subequations}
    \label{eq:trel}
    \begin{align}
        s_A(n) & = \theta+(1+\delta)t_D(n)+d_D(n)/c+z_A(n), \\
        s_D(n) & = \theta+(1+\delta)t_A(n)-d_A(n)/c+z_D(n), 
    \end{align}
\end{subequations}
where $d_D(n)$ and $d_A(n)$ are the inter-vehicle distances at times $t_D(n)$ and
$t_A(n)$, where $c$ is the speed of light, and where $z_A(n)$ and $z_D(n)$ are
additive receiver noise terms assumed to be \iid with mean zero and variance
$\sigma^2$. The timestamp relation~\eqref{eq:trel} assumes
that the vehicle movement and the clock drift are negligible during the packet time of
flight between the two vehicles. Since this time of flight is less than $300$ ns
for vehicle distances less than $100$ m, this assumption is reasonable.

\begin{figure}[htbp]
    \centering 
    \includegraphics{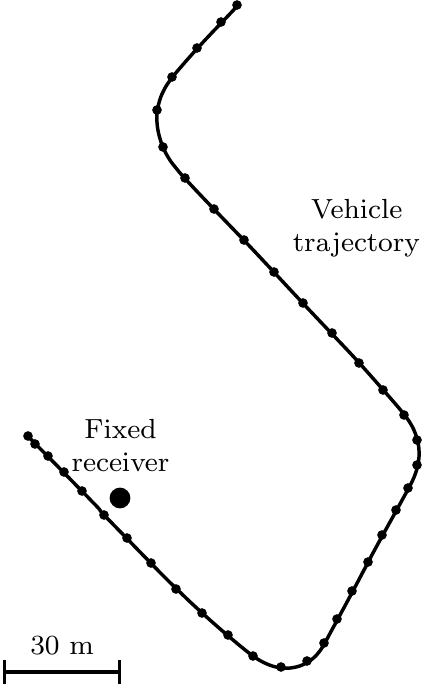} 
    \includegraphics{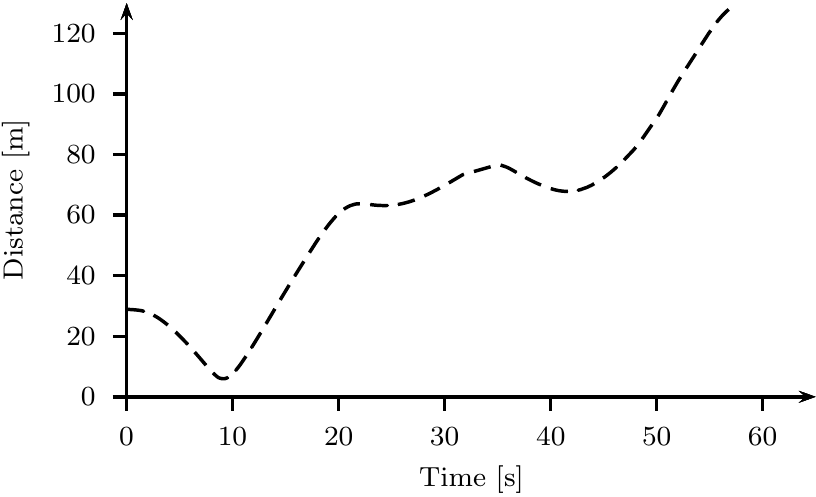} 

    \caption{Measured sample vehicle movement and corresponding distances to a
        fixed receiver (indicated by the large solid dot) as a function of time.}
    \label{fig:movement}
\end{figure}

The relation~\eqref{eq:trel} links the departure and arrival timestamps to the
inter-vehicle distances $d_D(n)$ and $d_A(n)$. Because of the movement of the
vehicles, these distances are time-varying (i.e., change as a function of $n$).
A typical vehicle trajectory and the corresponding distances of the vehicle to the fixed receiver are shown in
Fig.~\ref{fig:movement}. Since the vehicle distances can change by several tens
of meters per second, this time-varying nature of the distances needs again to
be taken into account when using periodic packet broadcasts for ranging. For example, in DSRC
with $T = 0.1$ s, the change in distances can be on the order of several meters
between successive message arrivals.

Recall from Fig.~\ref{fig:timeline} that at time $t_A(n)$ of the arrival of
message $n$ at the local vehicle, this vehicle has access to all its own
measured timestamps $t_D(i)$ and $t_A(i)$ with $i \leq n$. From the received
messages, it has also access to the remotely measured timestamps $s_A(i)$ with
$i\leq n$ and $s_D(i)$ with $i < n$ (note that, due to hardware limitations,
$s_D(n)$ is \emph{not} available at the local vehicle at time $t_A(n)$).  Using
this information, our goal in the remainder of this paper is to estimate at time
$t_A(n)$ the distance $d_A(n)$ between the vehicles at that time instant.

\begin{remark}[\emph{Comparison with RTT Ranging}]
    One standard approach to measure distances between unsynchronized
    communication devices is to use unicast RTT based ranging.  In this
    approach, a first communication device sends a unicast ranging request to a
    second communication device. The second communication device responds with a
    unicast acknowledgment. By measuring this round-trip time, the first device
    can then estimate the distance between the two devices.

    In the formalism of this paper, RTT ranging computes the distance estimate
    \begin{align}
        \label{eq:rtt}
        \hat{d}_A(n) 
        & = \tfrac{c}{2}\bigl((t_A(n)-t_D(n))-(s_D(n)-s_A(n))\bigr) \\
        & = \tfrac{1}{2}(d_A(n)+d_D(n)) 
        +\tfrac{c}{2}\bigl(\delta(t_D(n)-t_A(n))+z_A(n)-z_D(n)\bigr). \notag
    \end{align}
    In typical RTT applications, the time $t_A(n)-t_D(n)$ between transmitting
    the ranging request and receiving the corresponding acknowledgment is small.
    For example, in the RTT-based WiFi fine-timing measurement protocol, the
    delay between these two events is on the order of $0.01$ ms. Assuming a
    $\delta$ of $10$ ppm, the clock drift during that time is on the order of
    $0.1$ ns, resulting in a ranging error of approximately $1.5$ cm, which is
    negligible for most applications. Put differently,
    \begin{equation*}
        \tfrac{c}{2}\bigl(\delta(t_D(n)-t_A(n)) \approx 0.
    \end{equation*}
    At the same time, the vehicular movement during those $0.01$ ms is at most
    $0.1$ cm, so that the change in distances over this period can be ignored
    as well. Put differently,
    \begin{equation*}
        \tfrac{1}{2}(d_A(n)+d_D(n)) \approx d_A(n).
    \end{equation*}
    Hence, for unicast RTT ranging,
    \begin{equation*}
        \tfrac{c}{2}\bigl((t_A(n)-t_D(n))-(s_D(n)-s_A(n))\bigr) 
        \approx d_A(n)+\tfrac{c}{2}\bigl(z_A(n)-z_D(n)\bigr)
    \end{equation*}
    is an accurate estimate of the vehicular distances at time $t_A(n)$, and
    both drift and vehicular movement can be completely ignored.

    As already alluded to earlier, the situation changes completely for periodic
    broadcast ranging considered in this paper, where $T$ can be quite large.
    In the DSRC example, $T = 0.1$ s. As a result the clock drift error term
    \begin{equation*}
        \tfrac{c}{2}\bigl(\delta(t_D(n)-t_A(n))
    \end{equation*}
    can be on the order of $150$ m, and the vehicular movement error term
    \begin{equation*}
        d_A(n)-\tfrac{1}{2}(d_A(n)+d_D(n))
        = \tfrac{1}{2}(d_A(n)-d_D(n))
    \end{equation*}
    can be on the order of several meters. Hence, the simple RTT range
    estimator~\eqref{eq:rtt} fails in the periodic broadcast setting, and more
    sophisticated estimators, explicitly taking into account clock drift and
    vehicular movement, are needed. We present such an estimator in the next
    section.
\end{remark}

\section{Ranging Algorithm}
\label{sec:main}

We now describe an algorithm for ranging based on periodic broadcasts. The main
estimation algorithm is presented in Section~\ref{sec:main_poly}.
Section~\ref{sec:main_compression} proposes a compression scheme for the remote
timestamps. How these algorithms can be extended to handle arbitrary number of
vehicles is discussed in Section~\ref{sec:main_multiple}.

\subsection{Distance Estimation}
\label{sec:main_poly}

Recall from~\eqref{eq:clock} that the local time $t$ and the remote time $s$ are
related through a clock offset $\theta$ and a clock drift $\delta$. The clock
offset can be handled in a straightforward way by defining the timestamp
differences
\begin{align*}
    \dtd(n) & \defeq t_D(n)-t_D(n-1), \\
    \dsa(n) & \defeq s_A(n)-s_A(n-1), \\
    \dta(n) & \defeq t_D(n)-t_A(n-1), \\
    \dsd(n) & \defeq s_A(n)-s_D(n-1).
\end{align*}
Observe that these quantities were chosen such that
$\dtd(n)$, $\dsa(n)$, $\dta(n)$, and $\dsd(n)$ are all available
at the local vehicle at time $t_A(n)$ (see Fig.~\ref{fig:timeline} in
Section~\ref{sec:system}). 

The timestamp relation~\eqref{eq:trel} can then be rewritten as
\begin{subequations}
    \label{eq:dtrel}
    \begin{align}
        \dsa(n)
        & = (1+\delta)\dtd(n)+(d_D(n)-d_D(n-1))/c+\Delta z_A(n), \\
        \dsd(n) 
        & = (1+\delta)\dta(n)+(d_D(n)+d_A(n-1))/c+\Delta z_D(n), 
    \end{align}
\end{subequations}
with
\begin{subequations}
    \label{eq:dz}
    \begin{align}
        \Delta z_A(n) & \defeq z_A(n)-z_A(n-1), \\
        \Delta z_D(n) & \defeq z_A(n)-z_D(n-1).
    \end{align}
\end{subequations}
Thus, the transformed measurements are invariant with respect to the clock
offset $\theta$.  One could further transform the measurements to achieve
invariance with respect to the clock drift $\delta$ as well (see
Section~\ref{sec:main_compression}).  However, this would lead to significant
noise amplification, and we instead opt to explicitly estimate the nuisance
parameter $\delta$.

As was pointed out in Section~\ref{sec:system}, the clock drift $\delta$ can be
assumed to be constant during a window of a few seconds. From
Fig.~\ref{fig:movement} in Section~\ref{sec:system}, we see that during the same
timescale the inter-vehicular distances vary smoothly. Similar to
\cite{rajan2012joint}, we make a polynomial approximation of the variation of
these inter-vehicle distances as a function of time.  Due to the sharp peaks
occurring when two vehicles pass each other, we choose the degree of this local
polynomial approximation to be two. Formally, we are thus making the
approximation
\begin{subequations}
    \label{eq:poly}
    \begin{align}
        d_A(n) \approx a_0+a_1t_A(n)+a_2t_A^2(n), \\
        d_D(n) \approx a_0+a_1t_D(n)+a_2t_D^2(n).
    \end{align}
\end{subequations}

Fix a window size $w$, and assume both the clock drift and the distance
approximations are valid over $w$ time periods. From~\eqref{eq:dtrel}
and~\eqref{eq:poly}, the polynomial coefficients $a_0$, $a_1$, $a_2$, and the
nuisance parameter $\delta$ can then be estimated via least-squares as the
minimizer $(\hat{\delta}, \hat{a}_0, \hat{a}_1, \hat{a}_2)$ of
\begin{align*}
    & \bigl(c\dsa(n)-(1+\delta)c\dtd(n)
    -a_1t_D(n)-a_2t_D^2(n)+a_1t_D(n-1)+a_2t_D^2(n-1)\bigr)^2 \notag \\
    & \quad {} + \bigl(c\dsd(n)-(1+\delta)c\dta(n)
    -2a_0-a_1t_D(n)-a_2t_D^2(n)-a_1t_A(n-1)-a_2t_A^2(n-1)\bigr)^2 \notag \\
    & \quad {} + \dots \notag \\
    & \quad {} + \bigl(c\dsa(n-w+1)-(1+\delta)c\dtd(n-w+1)-\dots\bigr)^2 \notag \\
    & \quad {} + \bigl(c\dsd(n-w+1)-(1+\delta)c\dta(n-w+1)-\dots\bigr)^2
\end{align*}

This equation can be rewritten in matrix form as
\begin{equation*}
    \min_{\bm{x}}\, \norm{\bm{\beta}-\bm{B}\bm{x}}^2,
\end{equation*}
with 
\begin{align*}
    \bm{\beta} & \defeq c\cdot 
    \begin{pmatrix}
        \Delta\bm{s}_A(n-w+1:n)-\Delta\bm{t}_D(n-w+1:n) \\
        \Delta\bm{s}_D(n-w+1:n)-\Delta\bm{t}_A(n-w+1:n)
    \end{pmatrix} \\
    \bm{B} & \defeq 
    \begin{pmatrix}
        c\Delta\bm{t}_D(n-w+1:n) & \!\bm{0} & \!\bm{t}_D(n-w+1:n)-\bm{t}_D(n-w:n-1) & \!\bm{t}_D^2(n-w+1:n)-\bm{t}_D^2(n-w:n-1) \\
        c\Delta\bm{t}_A(n-w+1:n) & \!\bm{2} & \!\bm{t}_D(n-w+1:n)+\bm{t}_A(n-w:n-1) & \!\bm{t}_D^2(n-w+1:n)+\bm{t}_A^2(n-w:n-1)
    \end{pmatrix} \\
    \bm{x} & \defeq
    \begin{pmatrix}
        \delta & a_0 & a_1 & a_2
    \end{pmatrix}^\T
\end{align*}
and where we have used the notation
\begin{equation*}
    \bm{t}^2(n-w+1:n)
    \defeq
    \begin{pmatrix}
        t^2(n-w+1) & t^2(n-w+2) & \dots & t^2(n)
    \end{pmatrix}^\T.
\end{equation*}
The minimizer of this least-squares problem is
\begin{equation*}
    \hat{\bm{x}} = (\bm{B}^\T\bm{B})^{-1}\bm{B}^\T\bm{\beta},
\end{equation*}
from which we can recover the desired distance estimate
\begin{equation}
    \label{eq:dhat}
    \hat{d}_A(n) = \hat{a}_0+\hat{a}_1t_A(n)+\hat{a}_2t_A^2(n).
\end{equation}
As we will see next, this estimate can be improved upon by taking the correlated
nature of the transformed ranging noise into account. 

Recall from~\eqref{eq:dz} that the transformed noise $\Delta z_A(n)$ and $\Delta
z_D(n)$ is no longer white. In fact, setting
\begin{align*}
    \bm{z}_A & \defeq z_A(n-w:n) \\
    \bm{z}_D & \defeq z_D(n-w:n)
\end{align*}
and
\begin{align*}
    \Delta\bm{z}_A & \defeq \Delta z_A(n-w+1:n) \\
    \Delta\bm{z}_D & \defeq \Delta z_D(n-w+1:n),
\end{align*}
we can write
\begin{equation*}
    \begin{pmatrix}
        \Delta\bm{z}_A \\
        \Delta\bm{z}_D 
    \end{pmatrix}
    = \bm{J} 
    \begin{pmatrix}
        \bm{z}_A \\
        \bm{z}_D 
    \end{pmatrix}
\end{equation*}
with
\begin{equation*}
    \bm{J} \defeq
    \begin{pmatrix}
        (\bm{0}\ \bm{I})-(\bm{I}\ \bm{0}) & \bm{0} \\
        (\bm{0}\ \bm{I}) & -(\bm{I}\ \bm{0})
    \end{pmatrix}.
\end{equation*}
Hence,
\begin{equation*}
    \E\bigl( 
        (\Delta\bm{z}_A;\ \Delta\bm{z}_D)
        (\Delta\bm{z}_A^\T\ \Delta\bm{z}_D^\T)
    \bigr)
    = \sigma^2 \bm{J}\bm{J}^\T
    \defeq \bm{\Sigma}.
\end{equation*}

Thus, $\bm{\Sigma}^{-1/2}(\bm{\beta}-\bm{B}\bm{x})$ whitens the transformed ranging noise. The
corresponding whitened least-squares problem is
\begin{equation}
    \label{eq:min}
    \min_{\bm{x}}\, \norm{\bm{\Sigma}^{-1/2}\bm{\beta}-\bm{\Sigma}^{-1/2}\bm{B}\bm{x}}^2,
\end{equation}
with solution
\begin{equation*}
    \hat{\bm{x}} 
    = (\bm{B}^\T\bm{\Sigma}^{-1}\bm{B})^{-1}\bm{B}^\T\bm{\Sigma}^{-1}\bm{\beta}.
\end{equation*}
The estimate $\hat{d}_A(n)$ can be recovered from this as before (see
\eqref{eq:dhat}).

The computation of $\hat{d}_A(n)$ is performed at the local vehicle at each
time step $n$ to estimate the distance $d_A(n)$ from the remote vehicle
at the current time $t_A(n)$. We point out that all measurements required to
perform the computation of the estimate $\hat{d}_A(n)$ are available at the
local vehicle at time $t_A(n)$. 

\begin{figure}[htbp]
    \centering 
    \includegraphics{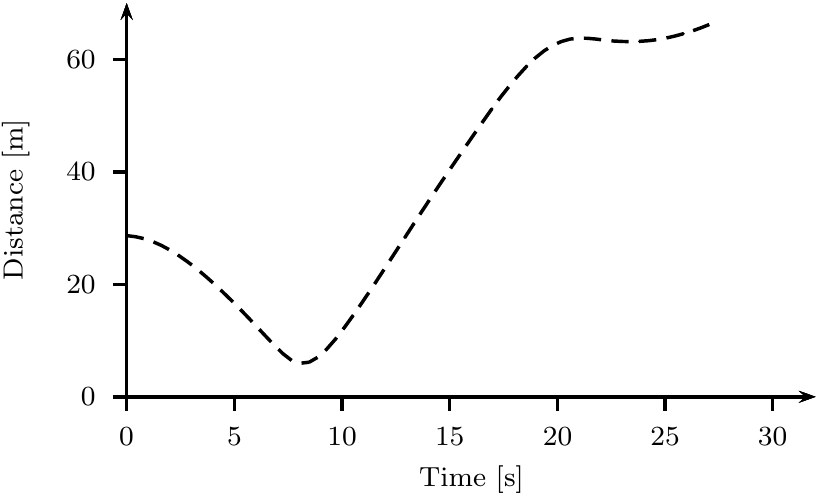} 
    \includegraphics{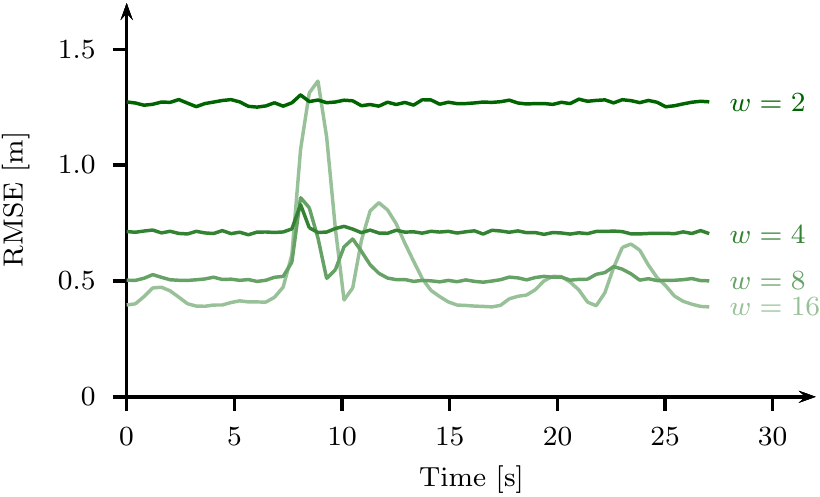} 

    \caption{Impact of window size on estimation error. The bottom figure plots
        the expected root mean-squared error (RMSE) (estimated using $10^4$
        Monte-Carlo simulations) as a function of time for several fixed window
        sizes. The top figure shows the corresponding actual distances.}
    \label{fig:window}
\end{figure}

Since we are estimating four parameters, and since there are two measurements at
each time step, the window size $w$ needs to be at least two for the matrix
$\bm{B}^\T\bm{\Sigma}^{-1}\bm{B}$ to be invertible. Larger values of $w$ result
in larger noise suppression. On the other hand, too large a value of $w$ will
lead to biased estimates due to model mismatch (see Fig.~\ref{fig:window}). For
the setting considered in this paper, a window size of $w = 8$ is a reasonable
choice.

\begin{figure}[htbp]
    \centering 
    \includegraphics{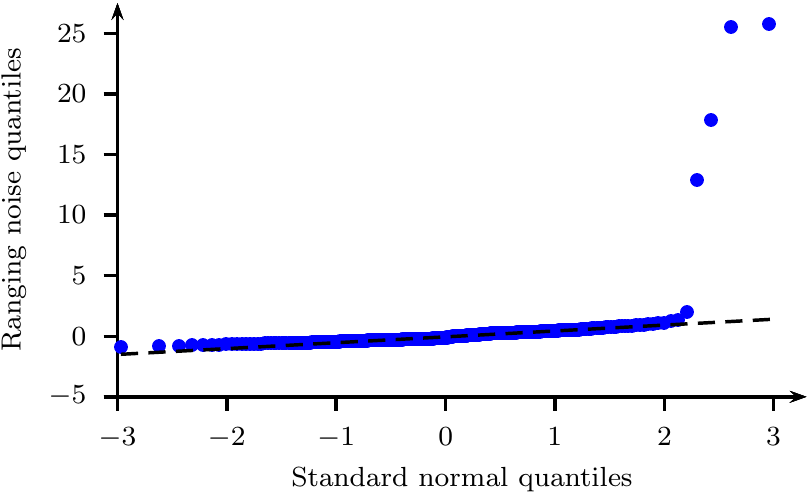} 

    \caption{Q-Q plot of ranging noise. The figure shows that the ranging noise
    exhibits heavy right tail caused by non-line-of-sight measurements.}
    \label{fig:qqplot}
\end{figure}

By performing experiments with known vehicle distances, the ranging noise terms
$z_D(n)$ and $z_A(n)$ can be directly measured. Fig.~\ref{fig:qqplot} shows the
Q-Q plot for the ranging noise, comparing the empirical ranging noise quantiles
to the quantiles of a standard normal distribution. For Gaussian ranging noise,
the points in the Q-Q plot should be linearly related. 

Fig.~\ref{fig:qqplot} indicates that the ranging noise is well modeled by a
Gaussian random variable except for the right tail, where the ranging noise
exhibits significant outliers. These outliers are caused by non-line-of-sight
(NLOS) range measurements, in which the direct path between the transmitter and
the receiver is obstructed, and the signal is instead observed via a longer path
containing reflections. Dealing with these large NLOS outliers is hence
critical, as will be discussed next.

Local polynomial regression can be made more robust to such NLOS outliers by
carrying out the least-squares procedure several times and reducing the weight
of those observations with large residuals~\cite{cleveland79}. Formally, denote
by 
\begin{equation*}
    \bm{e} \defeq \bm{\Sigma}^{-1/2}\bm{\beta}-\bm{\Sigma}^{-1/2}\bm{B}\hat{\bm{x}}
\end{equation*}
the vector of residuals in~\eqref{eq:min}. Compute the median $m$ of $\abs{\bm{e}}$
and the robustness weights
\begin{equation*}
    \gamma_i \defeq \bigl(1-e_i^2/(36s^2)\bigr)^2.
\end{equation*}
Then recompute a weighted least-squares solution with weights $\gamma_i$.
This procedure is repeated a few times. For our setting, five repetitions 
yielded good performance.

\subsection{Timestamp Compression}
\label{sec:main_compression}

Recall from Fig.~\ref{fig:timeline} in Section~\ref{sec:system} that the
departure and arrival timestamps $s_D(n)$ and $s_A(n)$ are transmitted from the
remote to the local vehicle and similarly for the timestamps $t_D(n)$ and
$t_A(n)$ from the local to the remote vehicle.  While for ranging between two
vehicles the number of bits that need to be transmitted is manageable, this may
no longer be the case when distances need to be estimated frequently and between
many vehicles. 

As an example, consider again the vehicular DSRC standard (which is a variant of
WiFi as mentioned before). The periodic broadcast messages are called basic
safety messages (BSM) in the DSRC standard. Each such BSM has a size of $142$
Bytes (in the simplest form) \cite{ansari2013}. Now, consider a scenario with
$20$ cars within radio range of each other, whose relative distances we aim to
estimate. A scenario of this size could easily arise in a multi-lane highway
setting, for example. Each vehicle then needs to piggyback $20$ timestamps on
each of its broadcast messages (one time of departure from its previously
broadcast message, and $19$ times of arrival of the broadcast messages of the
other cars, see also the discussion in Section~\ref{sec:main_multiple}). The FTM
protocol of the WiFi standard allocates $48$ bits to the transmission of each of
these arrival and departure timestamps quantized to a resolution of $0.1$ ns
\cite{11mc}. This results in a total of $120$ Bytes of data that need to be
piggybacked on the broadcast message.  Thus, the timestamp overhead to enable
the ranging is comparable to the size of the original BSM packet itself. 

In this section, we discuss a method to compress the measured timestamps to
reduce the overhead of communicating them. The proposed method achieves this
compression by making use of prior positional information. Specifically, by
using previously received and measured timestamps and by assuming that the
vehicles are moving at some bounded speed.

The compression of the timestamps is performed by discarding their
most-significant bits. The intuition is that since each vehicle is getting
periodic ranging and position estimates from the neighboring vehicle, an
approximate distance estimate at the current time is known. This approximate
distance information allows to reconstruct the most-significant timestamp bits.
It is only the lower-order bits that are needed to enhance the estimation
accuracy, and which consequently need to be transmitted. 

Formally, denote by $[\dsa(n)]$ the compressed timestamp corresponding to
$\dsa(n)$. Then
\begin{equation*}
    [\dsa(n)] \defeq 10^{10}\cdot\dsa(n) \bmod 2^L,
\end{equation*}
for some fixed positive integer $L$ to be specified later. Assuming that
$\dsa(n)$ is measured in integer multiples of $0.1$ nanoseconds, then the
quantity $[\dsa(n)]$ can be encoded using $L$ bits. These $L$ bits are then
transmitted from the remote to the local vehicle. 

To recover $\dsa(n)$ from $[\dsa(n)]$, observe that
\begin{equation*}
    10^{10}\cdot\dsa(n) = [\dsa(n)]+k_A(n) 2^L
\end{equation*} 
for some integer $k_A(n)$. Thus, the aim of the local vehicle is to find
$k_A(n)$. To this end, we make use of the fact that, by~\eqref{eq:dtrel},
\begin{equation}
    \label{eq:ratio}
    \frac{\dsa(n)}{\dtd(n)} 
    = 1+\delta+\frac{d_D(n)-d_D(n-1)}{c\dtd(n)}+\frac{\Delta z_A(n)}{\dtd(n)}
\end{equation}
so that
\begin{align*}
    \frac{\dsa(n)}{\dtd(n)} & -\frac{\dsa(n-1)}{\dtd(n-1)} \\
    & = \frac{d_D(n)-d_D(n-1)}{c\dtd(n)}-\frac{d_D(n-1)-d_D(n-2)}{c\dtd(n-1)} 
    +\frac{\Delta z_A(n)}{\dtd(n)} -\frac{\Delta z_A(n-1)}{\dtd(n-1)} \\
    & \approx \frac{d_D(n)-2d_D(n-1)+d_D(n-2)}{cT} 
    +\frac{\Delta z_A(n)-\Delta z_A(n-1)}{T},
\end{align*}
where the approximation is that $\dtd(n)\approx\dtd(n-1)\approx T$, i.e., that
the intervals between broadcast transmissions are approximately constant.  If
the relative speed of vehicle movement is bounded, and having probabilistic
bounds on the behavior of the ranging noise, this implies that
\begin{equation}
    \label{eq:ratiobound}
    \bigg\lvert \frac{\dsa(n)}{\dtd(n)} - \frac{\dsa(n-1)}{\dtd(n-1)}\bigg\rvert \leq \rho
\end{equation}
with high probability for some positive constant $\rho$.  For a concrete example,
in the DSRC context, $T = 0.1$ s and we can assume a relative vehicle speed of
at most $100$ m/s. Further, assume a ranging noise $\Delta z_A(n)$ of less then
$10 \text{ m}/c$, corresponding to a ranging error on the order of $10$ m. Then,
$\rho = 300 \text{ m/s}/c$ (a unitless quantity).

Assume for the moment that $k_A(n-1)$ and hence also $\dsa(n-1)$ are known, and
we want to recover $k_A(n)$. This can be achieved by finding the value of
$\hat{k}_A(n)\in\Z$ such that
\begin{equation*}
    \frac{[\dsa(n)]+\hat{k}_A(n)2^L}{10^{10}\cdot\dtd(n)} \approx \frac{\dsa(n-1)}{\dtd(n-1)}.
\end{equation*}
Or, put differently,
\begin{equation*}
    \hat{k}_A(n) 
    \defeq \bigg\lfloor 
    \Bigl( 10^{10}\frac{\dtd(n)}{\dtd(n-1)}\dsa(n-1)-[\dsa(n)]\Bigr)/2^L 
    \bigg\rceil,
\end{equation*}
where the operator $\lfloor\cdot\rceil$ rounds its argument to the closest
integer. Observe that $\dtd(n)$ and $\dtd(n-1)$ are available at the local
vehicle since they are locally measured and $\dsa(n-1)$ is available by
assumption.

From~\eqref{eq:ratiobound} we see that if
\begin{equation*}
    \frac{2^L}{10^{10}\cdot\dtd(n)} > 2\rho,
\end{equation*}
then $\hat{k}_A(n)$ is equal to $k_A(n)$, and we can recover $\dsa(n)$ from its
compressed value $[\dsa(n)]$. The number of bits needed for the compression is
thus
\begin{equation*}
    L > \log\bigl(2\cdot 10^{10}\rho\cdot\dtd(n)\bigr).
\end{equation*}

In the DSRC setting, this becomes $L > 10.97$ showing that $11$ bits will be
sufficient to transmit $[\dsa(n)]$. Thus, instead of transmitting $48$ bits as
in the FTM standard, we only need $11$ bits with the proposed compression
method. In our experiments, we choose a slightly more conservative number of $L
= 15$ bits to handle situations with dropped packets.

So far, we have seen how to recover $\dsa(n)$ from $[\dsa(n)]$ assuming that
$\dsa(n-1)$ has already been recovered. To start the recovery process, the
initial measurement $\dsa(1)$ could be transmitted uncompressed. Alternatively,
the first two compressed values $[\dsa(1)]$ and $[\dsa(2)]$ can be decompressed
jointly by solving a Diophantine approximation problem as is explained next.

From~\eqref{eq:ratio}, we can bound the ratio
\begin{equation}
    \label{eq:ratiobound2}
    \bigg\lvert \frac{\dsa(n)}{\dtd(n)}-1 \bigg\rvert \leq \tilde{\rho}
\end{equation}
with high probability for some positive constant $\tilde{\rho}$ in a manner
similar to~\eqref{eq:ratiobound}.  Note that typically $\rho$
in~\eqref{eq:ratiobound} will be smaller than $\tilde{\rho}$
in~\eqref{eq:ratiobound2}, since the latter does not remove the dependence on
the unknown clock drift $\delta$. This upper bound limits the set of possible
values for $k_A(n)$ to
\begin{equation*}
    \mc{K}_A(n) \defeq
    \biggl\{ k\in\Z: 
        \bigg\lvert \frac{[\dsa(n)]+k2^L}{10^{10}\cdot\dtd(n)}-1 \bigg\rvert 
        \leq \tilde{\rho} 
    \biggr\}.
\end{equation*}
We can use this to jointly decompress the first two compressed timestamp
$[\dsa(1)]$ and $[\dsa(2)]$ by solving the Diophantine approximation problem
\begin{align*}
    \bigl( & \hat{k}_A(1), \hat{k}_A(2)\bigr) 
    \defeq \argmin_{\substack{k(1)\in\mc{K}_A(1) \\ k(2)\in\mc{K}_A(2)}}
    \bigg\lvert 
    \frac{[\dsa(1)]+k(1)2^L}{10^{10}\cdot\dtd(1)}
    -\frac{[\dsa(2)]+k(2)2^L}{10^{10}\cdot\dtd(2)} 
    \bigg\rvert.
\end{align*}

The compression and decompression of the other timestamps $\dsd(n)$, $\dta(n)$,
and $\dtd(n)$ is performed analogously and is not repeated here.

\subsection{Arbitrary Number of Vehicles}
\label{sec:main_multiple}

The derivations so far have considered only two vehicles, one local and one remote.
We now describe how to generalize to arbitrary number of vehicles, one local
vehicle (from the point of which we are describing the execution of the
estimation algorithm) and $K$ remote vehicles. As before, each vehicle sends
periodic broadcast messages with approximate period $T$. 

\begin{figure}[htbp]
    \centering 
    \includegraphics{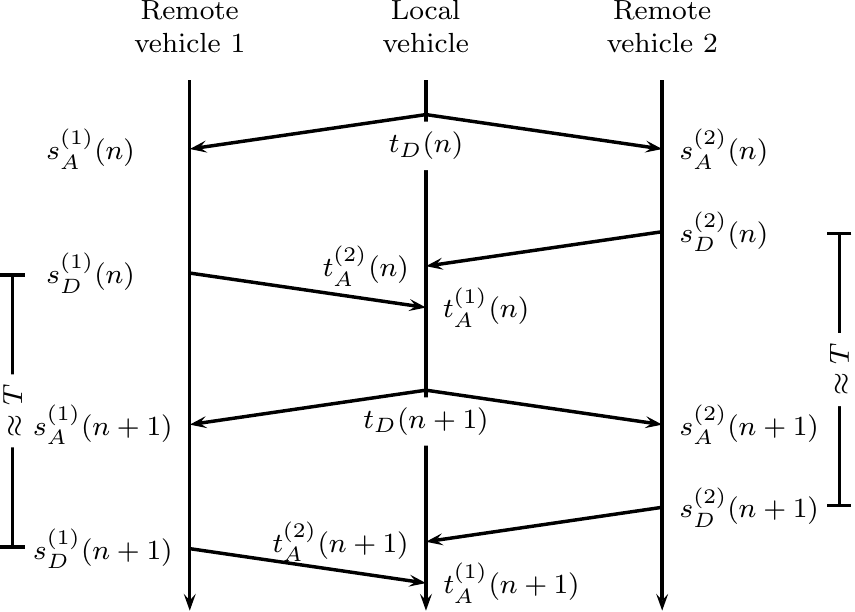} 

    \caption{Exchange of periodic broadcast messages for three vehicles. To simplify the figure,
    arrivals of the messages broadcast by the remote vehicles are only shown at
    the local vehicle (e.g., the arrival of the message broadcast by the second remote
    vehicle at time $s_D^{(2)}(n)$ at the first remote vehicle is not shown).}
    \label{fig:timeline4}
\end{figure}

The corresponding departure and arrival timestamps (each expressed with respect
to the clock of the vehicle where the event takes place) are depicted in
Fig.~\ref{fig:timeline4}. We denote by $t_D(n)$, the departure timestamps for
the local vehicle of broadcast message $n$ and by $s_D^{(k)}(n)$ with
$k\in\{1,2,\dots,K\}$ the departure timestamps at the $K$ remote vehicles of
broadcast message $n$. Since all transmissions are broadcast, each message
generates $K$ arrival timestamps, one at each other vehicle. We are interested
in the execution of the estimation algorithm at the local vehicle and to keep
notation manageable, we only consider the arrivals either observed at the local
vehicle or originating from the local vehicle. Denote by $t_A^{(k)}(n)$ the
arrival timestamp at the local vehicle of message $n$ broadcast from the remote
vehicle $k$. Similarly, denote by $s_A^{(k)}(n)$ the arrival timestamp at
remote vehicle $k$ of message $n$ broadcast from the local vehicle.

These timestamps are piggybacked onto the broadcast messages. For example, the
broadcast message sent by the local vehicle at time $t_D(n+1)$ carries the
(compressed) timestamp values of $t_D(n)$ and $t_A^{(k)}(n)$ for
$k\in\{1,2,\dots,K\}$ together with an ID identifying the car to which the time
stamp corresponds. Thus, every broadcast message contains $K+1$ piggybacked time
stamps and car IDs.

From the received piggybacked timestamps, the local vehicle estimates the
distances to all the $K$ remote vehicles by executing $K$ times the two-vehicle
range estimation algorithm described in Section~\ref{sec:main_poly}. In
particular, to estimate the distance to remote vehicle $k$, the estimation
algorithm is used with local timestamps $t_D(n)$, $t_A^{(k)}(n)$ and remote
timestamps $s_D^{(k)}(n)$, $s_A^{(k)}(n)$ with $n\in\{1,2,\dots\}$ (and for fixed
$k$).

\section{Experimental Evaluation}
\label{sec:experiments}

\subsection{Experimental Setup}
\label{sec:experiments_setup}

\begin{figure}[htbp]
    \centering 
    \includegraphics[scale=0.4, trim=3cm 0cm 2cm 1cm, clip=true]{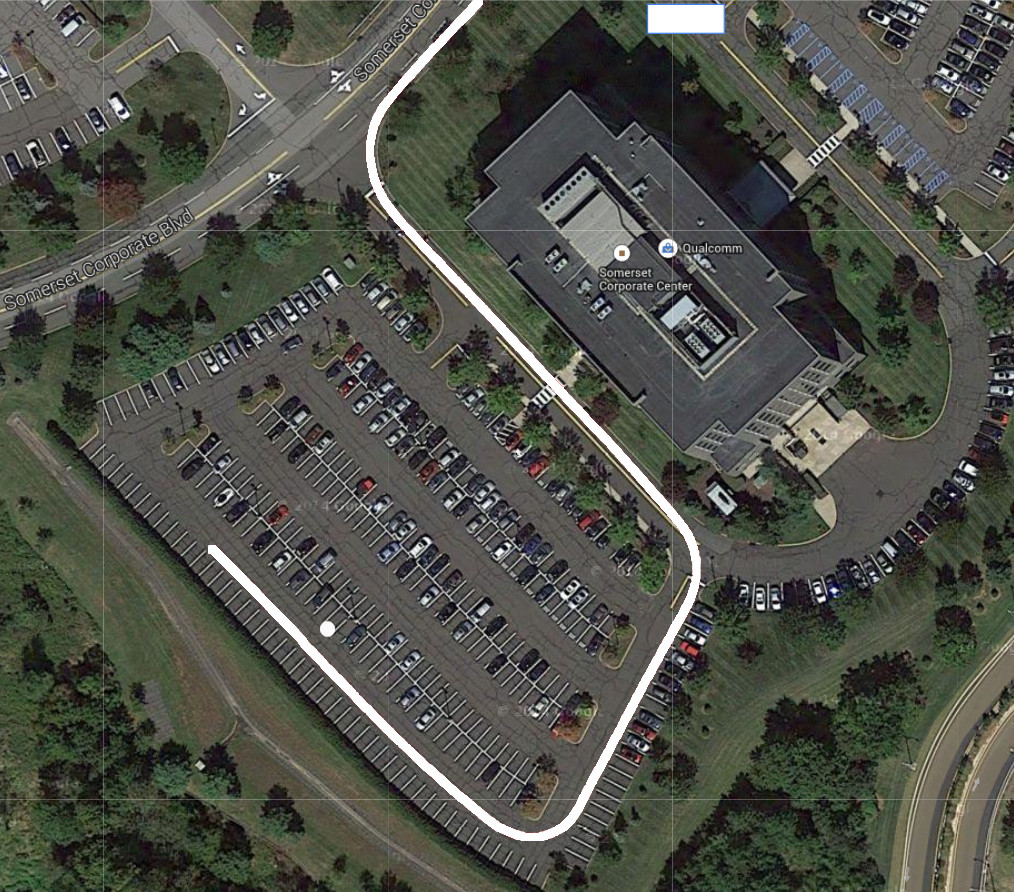} 

    \caption{Vehicle trajectory (directed from left to top) plotted on Google
    Maps. The white dot indicates the position of the static MDP.}
    \label{fig:gmapTraj}
\end{figure}

Experiments were carried out using Qualcomm-proprietary WiFi hardware for
emulating periodic wireless broadcasts. The WiFi hardware used consists of
mobile development platform (MDP) boards that have the capability to transmit
and receive packets in the WiFi bands. All experiments were carried out in the
parking lot at the Qualcomm New Jersey office (see Fig. \ref{fig:gmapTraj}). One
MDP was placed at a static location in the parking lot with the antenna placed
on an approximately $2$ m high pole. The second board was placed in a car with
the antenna placed on top of the vehicle. The antennas connected to both boards
were sharkfin antennas that are typically used in vehicles. The restriction of
using one static MDP was due to the availability of a single high-quality
ground-truth device.

\begin{figure}[htbp]
    \centering 
    \includegraphics{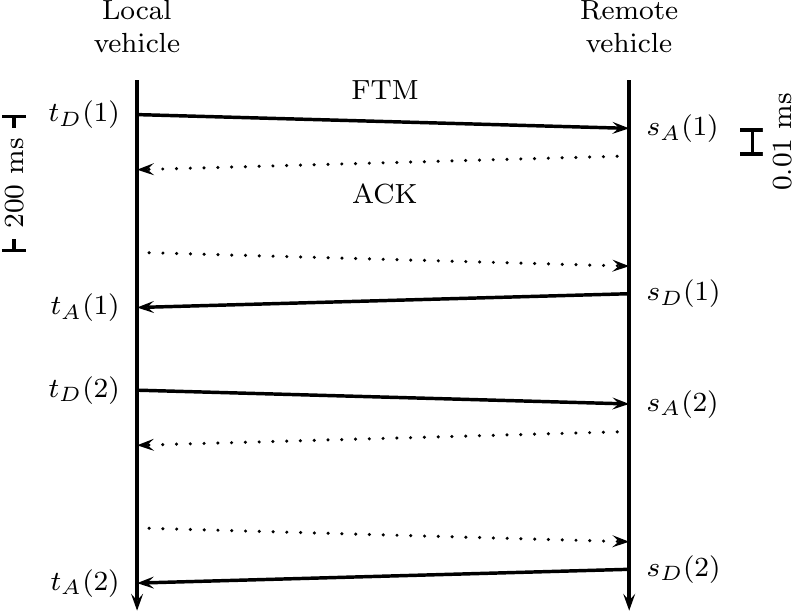} 
    \caption{RTT ranging message sequence representing the packets used for 
        evaluating the broadcast algorithm. Compare to Fig.~\ref{fig:timeline}
        in Section~\ref{sec:system}.}
    \label{fig:ftmdsrc}
\end{figure}

The MDP WiFi boards have the capability to transmit and receive ranging packets
for RTT ranging as specified in the 802.11mc standard \cite{11mc}. In our
experimental setup, ranging request packets were transmitted in the $5.2$ GHz
WiFi band from the MDP in the car and received by the static MDP. The static MDP
responds with an ACK to the request packets. The packet transmission rate was
kept at five per second (see Fig. \ref{fig:ftmdsrc}). The time-of-arrival and
time-of-departure timestamps were recorded at both the transmitter and receiver
and these were used for offline processing. 

To emulate broadcast ranging, the timestamps of every alternate ranging request
packet and every alternate ACK packet were used for processing. An example is
shown in Fig.~\ref{fig:ftmdsrc}. The FTM packets transmitted by the MDP in the
car (local vehicle) can be thought of as a proxy for the broadcast DSRC packets
from this car and the alternate ACK responses from the static MDP (remote
vehicle) can be taken as a proxy for its own broadcast DSRC packets. These are
denoted by the solid lines in the figure. Note that the time delay between the
FTM request and the alternate ACK's is on the order of $200$ ms. Thus the
effects of clock drifts will be prominently seen in the timestamps and need to
be accounted for.  Further, this method also helps us to compute distance
estimates using RTT based ranging for a baseline comparison by using the ACKs
that are immediately sent for every FTM request. 

All measurements were carried out at a bandwidth of $80$ MHz which is the
largest WiFi band supported by the MDP device. The received signal is then
filtered to different lower bandwidths to compare the ranging accuracy as a
function of the bandwidth.

The ground-truth device consists of a Novatel GPS module that has a
specification of decimeter accuracy in open sky environments. This device was
mounted on the vehicle after initially determining the ground truth of the
static MDP. The ground-truth device provides location fixes at a rate of $100$
per second. These are downsampled to match the rate of range estimates obtained
from the MDP. 

An important issue that needed to be resolved when using the ground-truth data
was that of the time synchronization between the estimates obtained from the
ground-truth device and those from the MDP. Note that the ground-truth device
provides measurements timestamped in both its own local clock and in GPS time.
The MDP devices however can only provide measurements timestamped in their local
clock. To address this synchronization issue, a Python script was used to query
the ground-truth device and the MDP board from a single laptop and timestamps
were attached to the measurements in this local clock. This provides
synchronization up to a few tens of milliseconds which is still comparable to
the time duration between the measurements. In the post-processing stage a
constant fixed delay between the ground truth and MDP measurements was therefore
manually adjusted to minimize the alignment error. 

The trajectory of the vehicle motion is shown in Fig.~\ref{fig:gmapTraj}.  The
driving environment is largely strong line-of-sight (LOS) except at certain
locations where the signal could be blocked by light poles or trees. There are
also reflections from the ground and other parked cars that can degrade the
estimation accuracy at lower bandwidths. We will see the effects of these in the
sections below.

\subsection{Comparison of Baseline RTT Scheme and Proposed PBR Scheme}
\label{sec:experiments_comparison}

\begin{figure}[htbp]
    \centering 
    \includegraphics{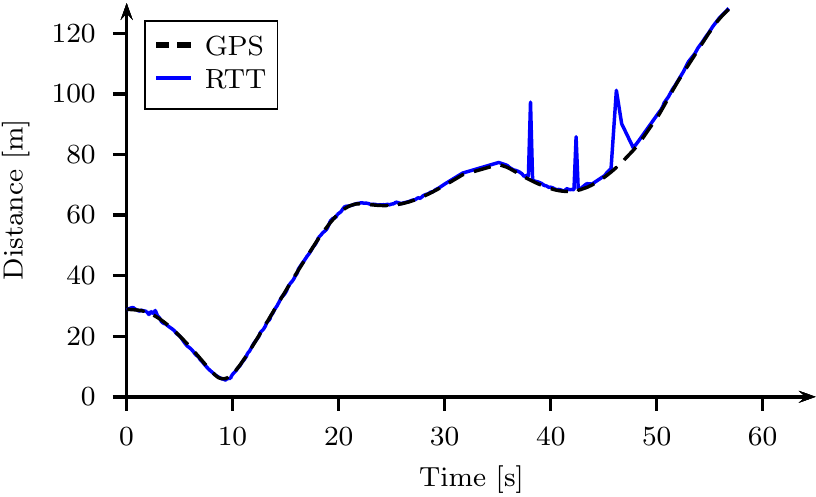} 
    \includegraphics{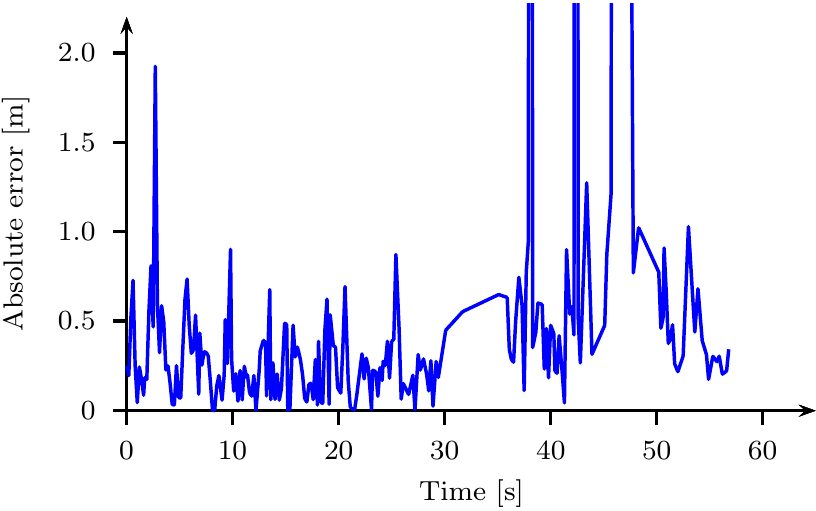} 

    \caption{Estimation error for the baseline unicast RTT-ranging approach ($80$ MHz ranging
    bandwidth).}
    \label{fig:unicast}
\end{figure}

A comparison of the performance of the baseline RTT scheme with the ground truth
at $80$ MHz bandwidth is shown in Fig. \ref{fig:unicast}. There is no averaging
or smoothing across time for this plot. One can see that the average errors
during the first $20$ s are well within half a meter. There are some large
errors mostly between the time instants $35$ s and $50$ s. These correspond to
the locations where the car was driving in front of the building as seen in Fig.
\ref{fig:gmapTraj}.  There are some light poles, trees and other parked cars
that contribute to multipath (between $40$ s and $50$ s) and packet loss
(between $30$ s and $35$ s). In particular, large multipath errors in this time
frame are potentially caused by signal reflection from the building adjacent to
the car trajectory.

\begin{figure}[htbp]
    \centering 
    \includegraphics{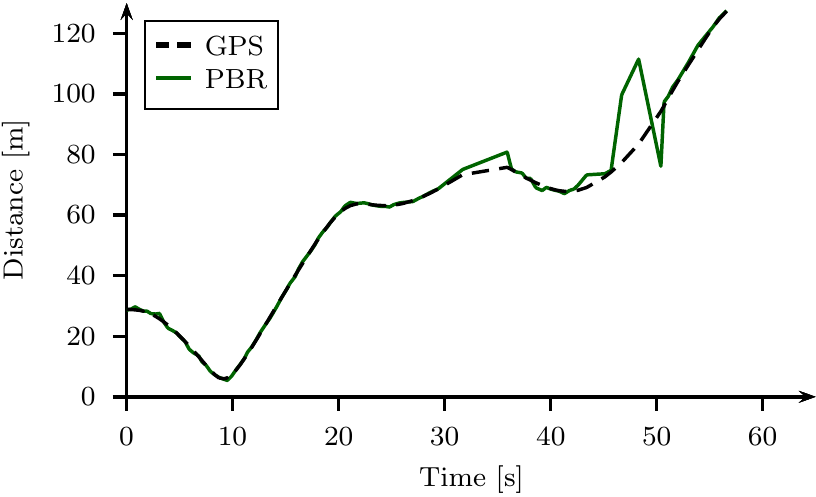} 
    \includegraphics{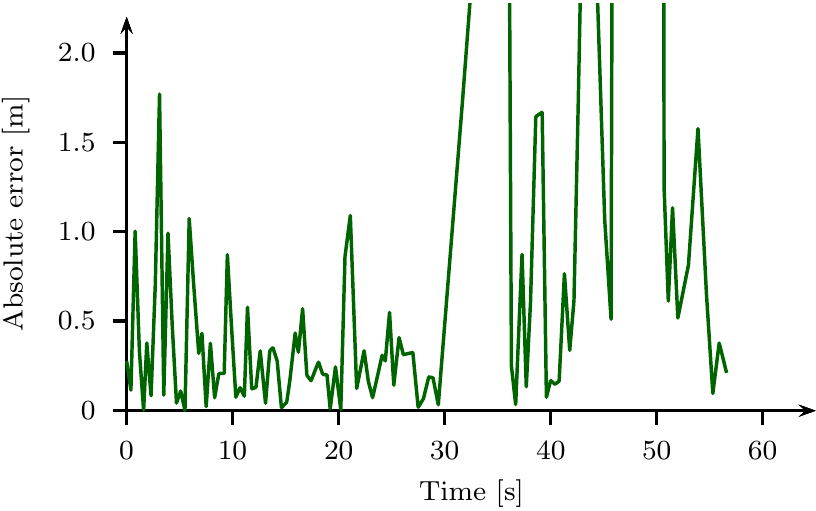} 

    \caption{Estimation error for proposed periodic broadcast ranging (PBR) approach
    ($80$ MHz ranging bandwidth).}
    \label{fig:broadcast80}
\end{figure}

Fig.~\ref{fig:broadcast80} shows a comparison of the errors between the proposed
periodic broadcast ranging (PBR) scheme and the ground truth. One can see that
the average errors during the first $20$ s are again well within half a meter.
Thus the proposed algorithm is able to estimate the clock drift fairly
accurately (recall that these can result in errors of more than $150$ m if not
taken care of). The plot looks smoother than RTT due to the local polynomial
approximation. Further this approach also eliminates some of the multipath
errors due to the local polynomial constraint wherein outliers are rejected. For
example, consider the multipath error in Fig.~\ref{fig:unicast} occurring around
time $38$ s. This is eliminated in the PBR scheme. However the other multipath
errors are not eliminated since these occur in bursts, particularly the one
close to $50$ s.  Further, the proposed scheme also has errors in certain cases
where there are packet losses and the local polynomial prediction can be
erroneous. This error is seen at time instants close to $35$ s during which the
prediction overshoots the actual trajectory.

\begin{figure}[htbp]
    \centering 
    \includegraphics{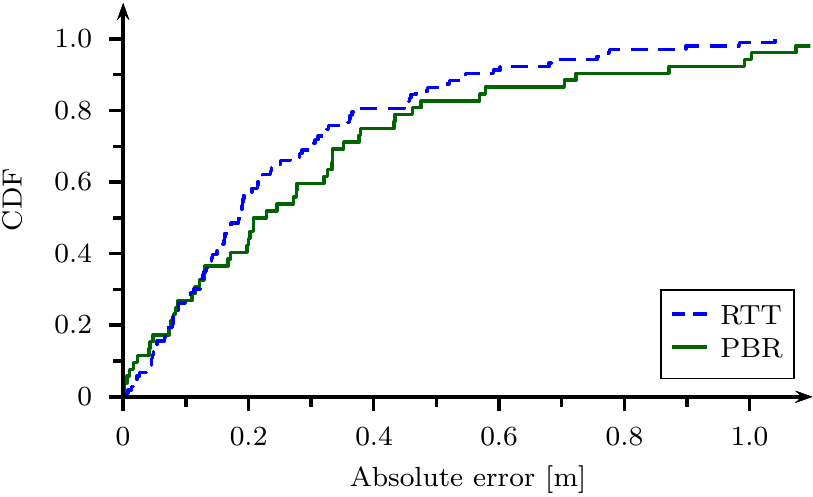} 

    \caption{Comparison of estimation error CDFs for unicast (RTT) and broadcast
        (PBR) approaches over the first $20$ s of vehicle trajectory ($80$ MHz ranging bandwidth).}
    \label{fig:cdf}
\end{figure}

Fig.~\ref{fig:cdf} shows the error cumulative distribution function (CDF)
between the RTT scheme and PBR for the first $20$ s of vehicle trajectory. There
is a slight degradation of around $5$ cm for PBR. However the overall average
error at the $50$ percentile is around $20$ cm and the $90$ percentile error is well
within $1$ m.

\subsection{Impact of Ranging Bandwidth}
\label{sec:experiments_bw}

\begin{figure}[htbp]
    \centering 
    \includegraphics{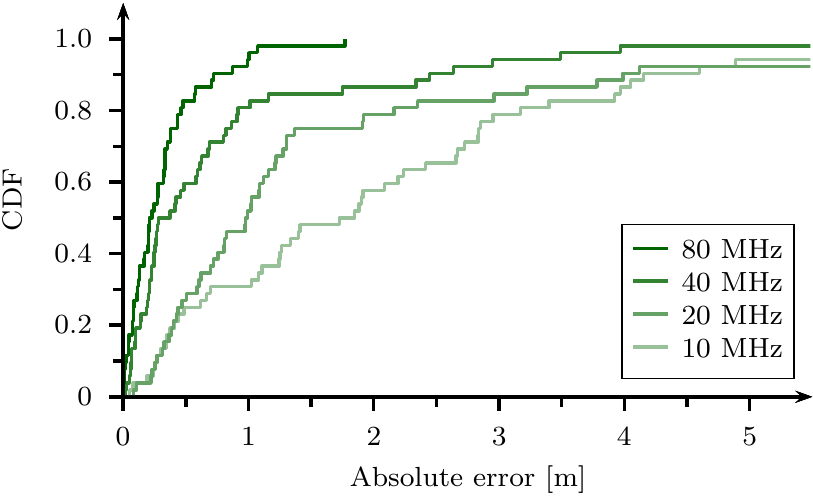} 

    \caption{Estimation error CDFs for PBR over the first $20$ s of vehicle
    trajectory for different ranging bandwidths.}
    \label{fig:bandwidth}
\end{figure}

Here we explore the effect of the transmission bandwidth on the ranging
accuracy.  The received $80$ MHz signal is an OFDM symbol with $256$ frequency
tones. This is downsampled to $128$, $64$, and $32$ tones to generate signals
occupying $40$ MHz, $20$ MHz, and $10$ MHz, respectively. The time-of-arrival is
then estimated based on these downsampled signals. Note that a factor of two
reduction in the bandwidth increases the pulse width in time by a factor of two,
thereby the accuracy is roughly expected to degrade by a factor of two
particularly when we have multipath. This effect is observed in our results as
well.  Fig.~\ref{fig:bandwidth} shows a comparison of the error CDFs for the
different bandwidths, and one can observe that the accuracy degrades by roughly
a factor of two (for example consider the $90$ percentile) with reducing
bandwidth.

\begin{figure}[htbp]
    \centering 
    \includegraphics{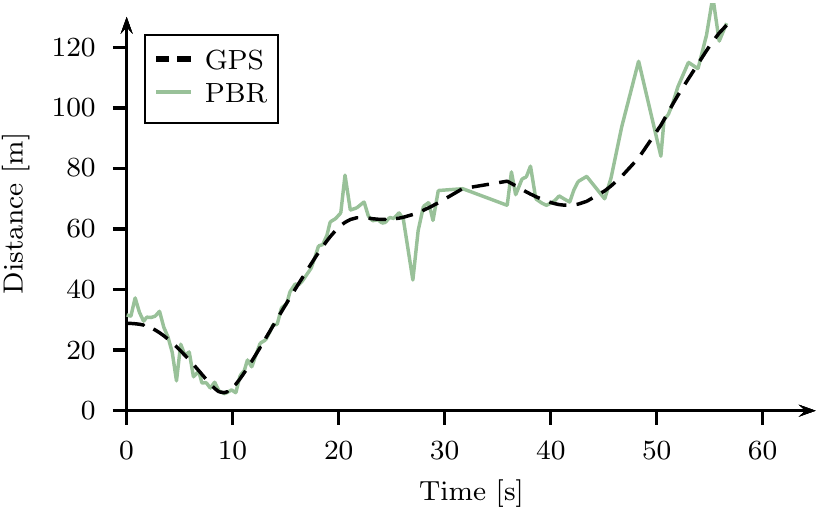} 
    \includegraphics{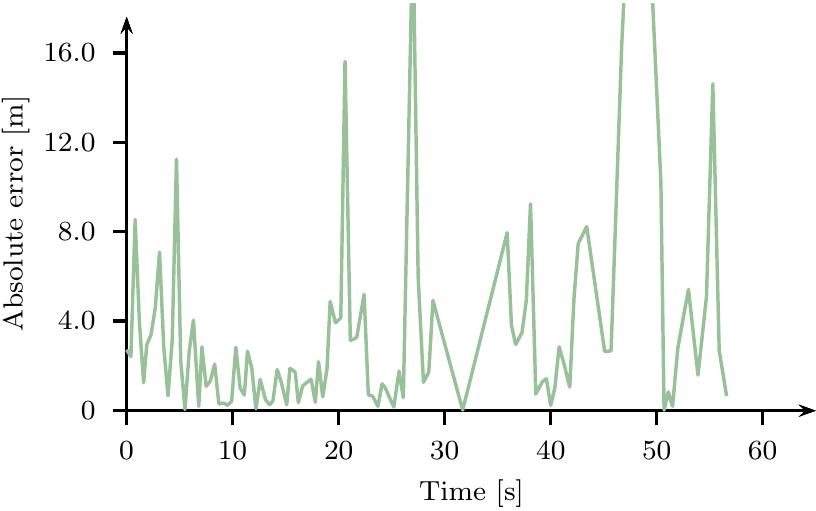} 

    \caption{Estimation error for proposed periodic broadcast ranging approach ($10$ MHz ranging
    bandwidth).}
    \label{fig:broadcast10}
\end{figure}

Fig.~\ref{fig:broadcast10} shows the ranging performance with $10$ MHz
transmission bandwidth, which is of interest in the DSRC context. During the
first $20$ s, the overall average error at the $50$ percentile is around $1.5$ m
as seen from the CDF curves.  In the linear section of the vehicle trajectory
between $10$ s and $20$ s, the errors are below $1$ m. In this case, there was
clear LOS between the two MDPs as opposed to the rest of the trajectory, during
which either other parked vehicles or obstructions like light poles resulted in
larger errors due to multipath.

\section{Conclusion}
\label{sec:conclusion}

In this work, we investigated the problem of range estimation between vehicles
using periodic broadcast transmissions. We proposed a broadcast ranging
algorithm that achieves accuracies comparable to unicast RTT based ranging.
Thus, the number of message exchanges needed for ranging can be reduced from
scaling quadratically in the number of vehicles for the unicast approach to only
linearly in the broadcast approach with only a small loss in ranging
performance.

Key challenges in broadcast ranging are the presence of clock offsets, clock
drifts, and vehicle motion that need to be accounted for.  The proposed
algorithm is able to efficiently handle these terms and to provide accurate
range estimates. We also propose a novel timestamp compression algorithm to
minimizes the packet overhead.

\section*{Acknowledgments}

The authors thank R.~Dugad, N.~Shah, S.~Yang, and L.~Zhang for their help in
collecting measurements.

\end{document}